\documentclass[10pt,letterpaper,twocolumn]{article} 

\usepackage{ol2}
\usepackage[draft,implicit=false]{hyperref}
\usepackage{amsmath}

\begin{document}

\twocolumn[ 

\title{\emph{Two photon quantum interference in plasmonics - Theory and Applications}}


\author{S. Dutta Gupta$^{1,2,*}$ and G. S. Agarwal$^{2}$}

\address{
$^1$School of Physics, University of Hyderabad, Hyderabad-500046, India\\
$^2$Department of Physics, Oklahoma State University, Stillwater, Oklahoma 74078, USA
 \\
$^*$Corresponding author: sdghyderabad@gmail.com
}

\begin{abstract}
We report perfect two photon quantum interference with near-unity visibility in a resonant tunneling plasmonic structure in folded Kretschmann geometry. This is despite absorption-induced loss of unitarity in plasmonic systems. The effect is traced to perfect destructive interference between the squares of amplitude reflection and transmission coefficients. We further highlight yet another remarkable potential of coincidence measurements as a probe with better resolution as compared to standard spectroscopic techniques. The finer features show up in both angle resolved and frequency resolved studies.
\end{abstract}


 ] 
 
 In recent years single photons \cite{hom,gsa}, besides being of fundamental interest in quantum information protocols, are finding applications in diverse areas like quantum metrology \cite{dowling}, target detection \cite{target1}, quantum imaging \cite{qi1,qi2}, etc. One naturally seeks a chip-level realization of many of these concepts and devices. In view of the tremendous advancements in theoretical and technological aspects in recent times, plasmonic systems are possible candidates for such implementations. There has been a great deal of research in probing the potentials of plasmonics for single-photon based devices. The main disadvantage of plasmonics in this context stems from the unavoidable losses due to the presence of metal, which gets accentuated near the surface and localized plasmon resonances. Key concepts like unitarity are lost. Thus the initial attempts on fusing plasmonics with single photons was to check whether the quantum features survive the passage through plasmonic structures. Plasmon assisted transmission of polarization entangled photons through a sub-wavelength array of holes was demonstrated \cite{woerdman}. A similar effect in terms of energy-time entanglement was reported later \cite{zbinden}. The excitation of single plasmons was demonstrated in several experiments. Heralded single plasmons were excited using entangled photons generated by spontaneous parametric down-conversion \cite{sahin}. In yet another experiment a CdSe quantum dot was excited in the near vicinity of a silver nanowire to generate the single plasmons \cite{lukin}. The Hong-Ou-Mandel (HOM) dip \cite{hom} was achieved in both these experiments. A chip-level implementation of the HOM interference using a plasmonic circuit was realized very recently \cite{zwiller}.
\begin{figure}[htbp]
\center{\includegraphics[width=7.5cm]{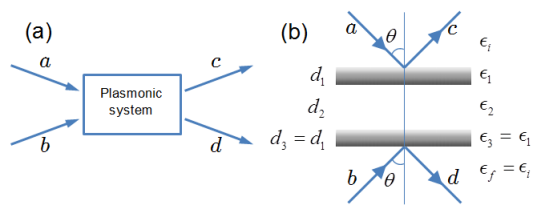}}
\caption{Schematic view of (a) the generic system and (b) the symmetric gap plasmon guide where two Kretschmann configurations are put together with a spacer gap.}
\label{fig:fig1}
\end{figure}
 \par
 In all the reported experiments on HOM interference in plasmonic systems, the dip never reaches the null level, since perfect unitarity is lost due to the presence of losses. In this letter we show that by a clever design of the system one can convert the disadvantage of plasmonics into a positive advantage. Indeed, one can realize perfect two photon quantum interference (TPQI) with visibility unity, when the coincidence count is reduced to the null. We also show that TPQI can be used as an efficient probe in plasmonic systems, since it offers a better resolution as compared to standard spectroscopic techniques. This is in contrast to super resolution studies of phase in optical interferometer \cite{dowling}. Here we show TPQI to be a better probe of a resonance which is characterized by losses. The complete destructive interference in perfect TPQI in a lossy system is somewhat reminiscent of recently reported coherent perfect absorption (CPA) where there is a near-total  suppression of scattering (simultaneous null transmission and null reflection) leading to interesting applications \cite{cpa1,cpa2,cpa3,sdgtf}. We present the necessary and sufficient conditions of having TPQI with an example of a plasmonic system of a folded Kretschmann configuration (KC) with a gap between the folded parts (otherwise known as a gap plasmon guide \cite{sdgtf}). In our system finite transmission is ensured through resonant tunneling \cite{pendry} via the excitation of the coupled surface plasmons. Note that simultaneous fulfillment of conditions for perfect TPQI and CPA is ruled out because of the conflicting phase relations for them. Further, we reveal yet another important feature of application potentials of TPQI. We propose two photon correlation as a spectroscopic probe with better resolution capabilities as compared to the standard attenuated total reflection (ATR) studies of intensity reflection or transmission. Features barely discernible in the latter can be captured in the two photon correlation spectra.
 \par
 Consider a generic situation shown in Fig.\ref{fig:fig1}a. Let $a$ and $b$ ($c$ and $d$) represent the Heisenberg operators for the input (output) quantum fields. The Heisenberg operators $a$, $b$, $c$ and $d$ and their adjoints satisfy the bosonic commutation relations. Since one excites the plasmon modes (both surface and bulk) in the plasmonic medium where the absorptionn is also important, the output fields are realted to the input fields via
 \begin{equation}
\label{eq1}
\begin{pmatrix}
 c \\
 d   
\end{pmatrix}
=S
\begin{pmatrix}
 a \\
 b   
\end{pmatrix}+
\begin{pmatrix}
 f_a \\
 f_b   
\end{pmatrix},~~~~~
S=
\begin{pmatrix}
 \alpha     &   \beta \\
\gamma     &  \delta
\end{pmatrix}
\end{equation}
where $f_a$ and $f_b$ represent the quantum noise sources so that the output fields satisfy commutation relations $[c,c^\dagger]=[d,d^\dagger]=1$, $[c,d^\dagger]=0$. The noise sources have the property that they do not contribute to the normally ordered correlations of the output field. The transformtion matrix $S$, which depends on the optical properties of the plasmonic material, is not unitary
\begin{equation}
\label{eq2}
S^\dagger S \ne 1,  ~~~~  SS^\dagger \ne 1.
\end{equation}
On the contrary when unitarity holds, the elements of $S$ satisfy the following
\begin{equation}
\label{eq3}
|\alpha|^2+|\gamma|^2=1,~~|\beta|^2+|\delta|^2=1,~~\alpha^*\beta+\gamma^*\delta=0.
\end{equation}
At the output the physical quantities that one can measure are the mean number of photons $I_c=<c^\dagger c>$, $I_d=<d^\dagger d>$ and the coincidence probability $C_q=<c^\dagger cd^\dagger d>$ for measuring one photon in each of the modes $c$ and $d$. The measured quantities will depend on the state of the input fields. We will consider two cases to highlight the importance of using single photons for plasmonic studies. We will assume that the input state is either a single photon state $|1,1>$ or a coherent state $|u,v>$. Using (\ref{eq1}) it is easy to show that 
\begin{eqnarray}
\label{eq4}
I_{cq}&=&|\alpha|^2+|\beta|^2, ~~ I_{cc}=|\alpha u+\beta v|^2, \nonumber\\
I_{dq}&=&|\delta|^2+|\gamma|^2, ~~ I_{dc}=|\gamma u+\delta v|^2, \nonumber\\
C_{q}&=&|\alpha \delta+\beta \gamma|^2,\nonumber \\
C_{c}&=&|\alpha u+\beta v|^2 |\gamma u+\delta v|^2=I_{cc}I_{dc}.
\end{eqnarray}
The subscripts $c$ and $q$ represent results with classical and quantum fields, respectively. The first subscripts in Eq.(4) refer to the field states. $C_c$ is the coincidence measurement results obtained with classical fields. We note that the vanishing of $I_{cc}$ and $I_{dc}$, and consquently $C_c$ gives the coherent perfect absorption, which crealy does not occur if there is no absorption. Further note that for coherent fields, the coincidence measurement gives no new information. For quantum fields and in particular for the case of single photons, $I_{cq}$ and $I_{dq}$ can never vanish. However, the coincidence measurement $C_q$ involves a coherent sum of two distinct contributions and can in principle be zero even in presence of absorption. We call this possibility as perfect TPQI. This is to differentiate it from the case when one uses two very weak coherent beams at single photon level. Although our discussion was in the context of a plasmonic system, our results are equally valid for other systems like optical resonators \cite{sahin2}, where resonant modes could be selectively excited.
\begin{figure}[t]
\center{\includegraphics[width=8cm]{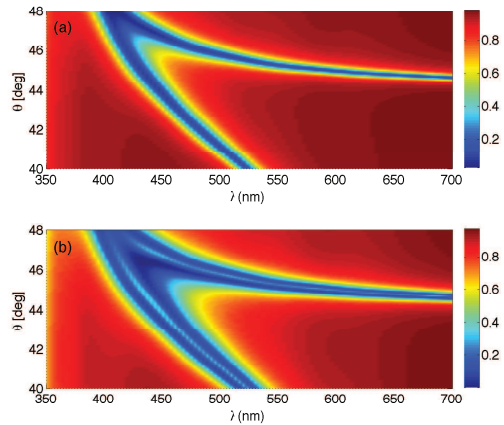}}
\caption{Dispersion diagram for the gap plasmon guide for (a) the  intensity reflection coefficient $R$  and (b) the quantum correlation parameter $C_q$ for $d_1=d_3=48~nm$, $d_2= 450~nm$, $\epsilon_i=\epsilon_f=2.28$, $\epsilon_2=1.0$.}
\label{fig:fig2}
\end{figure}
\par
Having discussed the general features we now analyze a specific symmetric structure shown in Fig.\ref{fig:fig1}b. Our prototype system consists of two folded Kretschmann configurations (KC) separated by a spacer gap. The motivation for choosing such a structure is to have resonant tunneling through the gap barrier via the excitation of the coupled plasmon modes. For sufficiently thin gap layer the plasmons of the top and bottom gap/metal interfaces can be coupled to lead to the symmetric and antisymmetric modes. These are not the same as the usual coupling in Sarid geometry through the metal film leading to long range and short range modes. Because of coupling through a non-lossy dielectric (gap), the decay rates are comparable and such structures are some times referred to as gap plasmon guides \cite{sdgtf}. Note that transmission through a standard KC is null since the waves are evanescent below the metal film when plasmons are excited. Let p-polarized signal and idler photons originating from a SPDC or any other single photon source be incident on the system from both sides as shown in figure. We further assume identical illumination $u=v$ for coherent input. For a symmetric structure $\alpha=\delta=r$, $\beta=\gamma=t$, where $r$ and $t$ are the amplitude reflection and transmission coefficients, respectively \cite{sdgreview}.  Then Eq.(\ref{eq4}) simplifies to
\begin{eqnarray}
\label{eq5}
I_{cq}& = & R+T=I_{dq}, \nonumber\\
C_q& = &|r^2+t^2|^2=(R-T)^2+U^2, \nonumber\\
C_c& = &|u|^2|r+t|^4=|u|^2(R+T+U)^2,
\end{eqnarray}
where 
\begin{equation}
\label{eq6}
U=r^*t+rt^*=2 \sqrt{RT} \cos(\Delta \phi),~~|r|^2=R,~~|t|^2=T.
\end{equation}
$R$ and $T$ are the intensity reflection and transmission coefficients, respectively. $\Delta \phi=\phi_r-\phi_t$ with $\phi_r$ and $\phi_t$ giving the phases of the amplitude reflection and transmission coefficients, respectively. The parameter U (hereafter referred to as the unitarity parameter) is nonzero for the plasmonic structure near resonant frequencies. It is clear from Eq.(5) that $C_q=0$ if $R=T,~ U=0$. For these conditions $C_c\ne0$ classical CPA is ruled out. The classical CPA occurs when $U\ne0$ and is equal to $-(R+T)$. We also note that in a nonabsorbing structure $U=0$, and for $R=T=1/2$ one has $C_q=0$. These are the conditions when HOM dip would occur.
\par
Eq.(\ref{eq5}) clearly justifies the choice of our resonance tunneling plasmonic structure, where the excitation of the plasmons are essential for having finite transmission and they are bound to leave their imprint on the two photon or two beam correlation spectra. We now show that quantum correlation spectra can offer a finer probe as compared to the usual attenuated total reflection (ATR) spectra of intensity reflection and transmission.
\begin{figure}[t]
\center{\includegraphics[width=7.8cm]{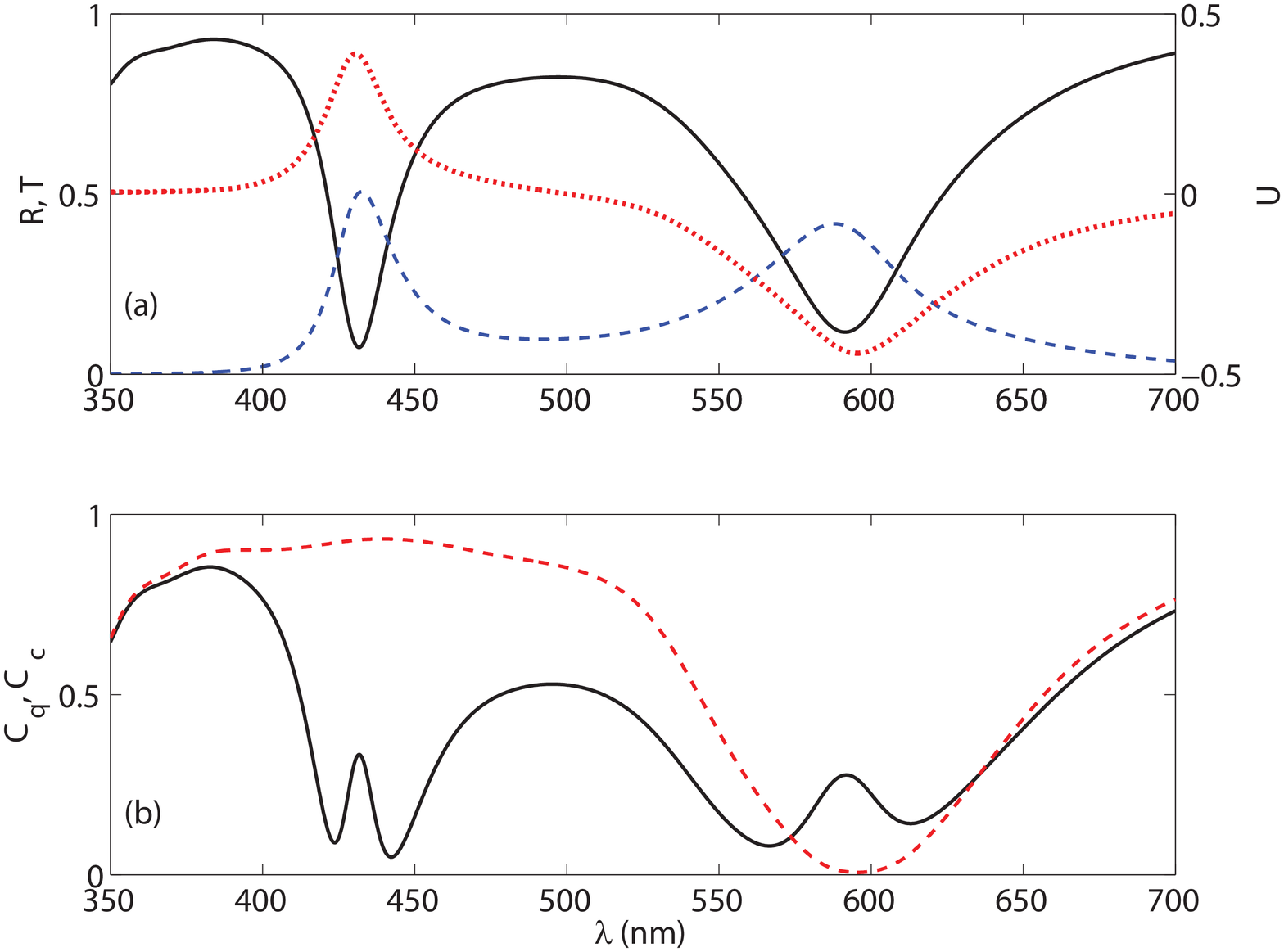}}
\caption{(a) Intensity reflection $R$ (solid curve), transmission $T$ (dashed curve) coefficients and unitarity parameter $U$ (dotted),  (b) $C_q$ (solid)  and $C_c$ (dashed-dot) as functions of wavelength $\lambda$ for $\theta=45^\circ$. Other parameters are as in Fig.~\ref{fig:fig2}.}
\label{fig:fig3}
\end{figure}
\begin{figure}[t]
\center{\includegraphics[width=7.8cm]{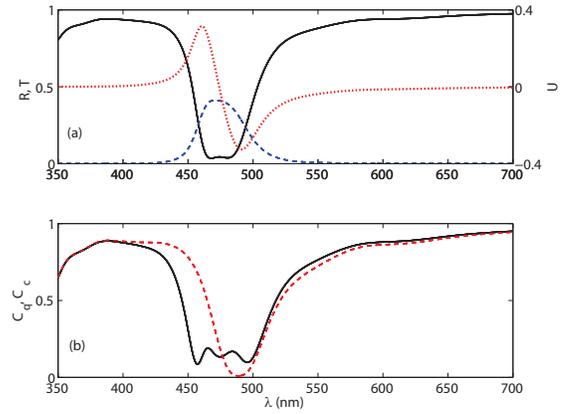}}
\caption{Same as in Fig.~\ref{fig:fig3} except that now $d_2=750~nm$. Other parameters are as in Fig.~\ref{fig:fig2}.}
\label{fig:fig4}
\end{figure}
\par
The results for intensity reflection coefficient $R$ and the quantum correlation parameter $C_q$ as functions of angle of incidence $\theta$ and wavelength $\lambda$ are shown in Figs. \ref{fig:fig2}a and \ref{fig:fig2}b. The following set of parameters were used for calculations 
 $d_1=d_3=48~nm$ , $d_2= 450~nm$, $\epsilon_i=1=\epsilon_f=2.28$, $\epsilon_2=1$, while gold was used for the metal films and $\epsilon_1=\epsilon_3$ for gold was taken from the work of Johnson and Christie \cite{johnson}. One can easily see the familiar splitting leading to the two branches due to the  symmetric and the antisymmetric coupled modes. However, the correlation spectra have more details and finer features enabling one to have a better resolution both in angle resolved or wavelength resolved studies. In order to show this more clearly we have plotted the relevant quantities in Figs. \ref{fig:fig3}a and \ref{fig:fig3}b for a specific angle of incidence $\theta=45^\circ$. Indeed the features in $C_q$ near split mode resonances are sharper due to the emergence of additional minima. The minima in $C_q$ occur when $R=T$ and they can be easily explained using Eq. (3). For example, for the leftmost minimum in Fig.\ref{fig:fig3}b, there is an increase in both $R-T$ and $U$ as one moves away from the minimum to the right. For movement to the left $R-T$ increases faster than the decease in $U$, leadining to an increase in $C_q$ from the minimum value. The emergence of sharper features in $C_q$ can be utilized to resolve better the emergent normal mode splitting for smaller coupling when the gap width $d_2$ is increased and when the responses in terms of $R$ and $T$ do not show the splittings clearly. One such case is shown in Fig. \ref{fig:fig4} for $d_2=750~nm$. One cannot infer about the split normal mode wavelengths by looking at  $R$ and $T$ in Fig.\ref{fig:fig4}a, while it can be easily done by inspecting $C_q$ (Fig.\ref{fig:fig4}b). 
 \par
 One can naturally enquire about how the quantum correlation results differ from the classical analogue. In fact, the expression for the classical correlation parameter $C_c$ is just the square of the total scattering on either side of the system  $|r+t|^2$ and a zero in the total scattering heralds the onset of CPA \cite{cpa2}. The minima in $C_c$ occur near the CPA condition $R+T+U\sim 0$ (see Figs. \ref{fig:fig3}b and \ref{fig:fig4}b). Needless to stress that the classical correlation is unable to capture the details of the split modes. 
 \par
\begin{figure}[t]
\center{\includegraphics[width=7.8cm]{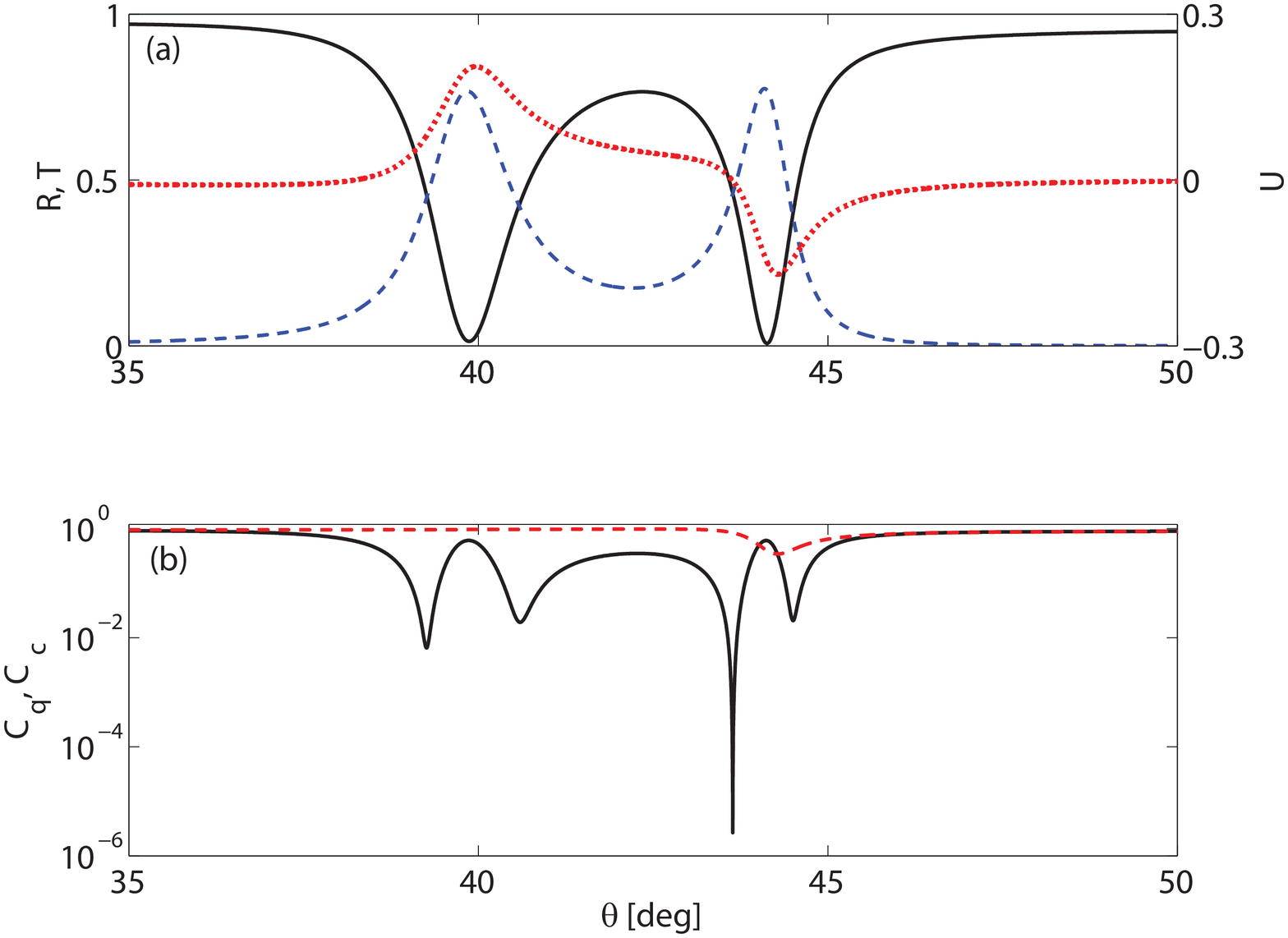}}
\caption{(a) $R$ (solid), $T$ (dashed) and $U$ (dotted), (b) $C_q$ (solid), and $C_c$ (dashed) as functions of  $\theta$ for $\lambda=650~nm$ and $d_1=30~nm$, $d_2=640.2~nm$. Other parameters are as in Fig.~\ref{fig:fig2}.}
\label{fig:fig5}
\end{figure}
\begin{figure}[h]
\center{\includegraphics[width=7cm]{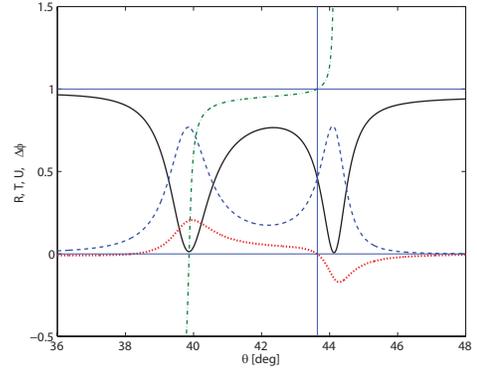}}
\caption{(a) $R$ (solid), $T$ (dashed), $U$ (dotted) and $\Delta\Phi$ (dash-dotted) in units of $\pi/2$ as functions of  $\theta$ for parameters of Fig. \ref{fig:fig3}. The vertical line depicts $\theta=43.64^\circ$ and the two horizontal lines show, respectively, $U=0$ and $\Delta\phi=1$. Other parameters are as in Fig.~\ref{fig:fig2}.}
\label{fig:fig6}
\end{figure}

 We now show that it is possible to have near-complete destructive interference (leading to $C_q \sim 0$) with the single photons in the lossy system. The complete cancellation with perfect TPQI is shown in Fig. \ref{fig:fig5}, where we considered thinner gold films with $d_1=d_3=30~nm$ $d_2=640.2~nm$ at $\lambda=650~nm$. In order to further demonstrate the flexibility of the system we have plotted the relevant quantities as functions of angle of incidence. As can be seen from \ref{fig:fig5}b, the quantum correlation parameter $C_q$ can drop to $10^{-6}$ at about $\theta=43.64^\circ$. In the context of classical correlation $C_c$, the system is far away from the CPA condition leading to a feeble dip. In order to have a deeper insight in the perfect TPQI we collected all the relevant physical quantities $R$, $T$, $U$, $\Delta \phi$ and show them in an expanded plot near $\theta=43.64^\circ$ in Fig. \ref{fig:fig6}. The vertical line marks the angle $\theta=43.64^\circ$, while the horizontal lines indicate $U=0$ and 
$\Delta \phi=\pi/2$. It is clear from Fig. \ref{fig:fig6} that the perfect TPQI occurs when the conditions $\Delta \phi=\pi/2$, $R=T\ne1/2$ and $U=0$ are simultaneously satisfied. In fact, $\Delta \phi=\pi/2$ implies $U=0$ for finite $R$ and $T$ (see Eq.(\ref{eq6})). In other words phase difference between $r^2$ and $t^2$ is $\pi$, enabling destructive interference in $C_q$. Further $R=T$ ensures complete cancellation leading to $C_q \sim 0$ implying perfect TPQI. Note that the TPQI dip occurs when $R=T$ and not at the split mode resonances. As mentioned earlier, in view of the inherent absorption due to metal films the perfect unitarity is lost and the perfect TPQI can be treated as a generalization of the HOM dip for lossy systems. In fact, unitarity is restored to the maximal extent ($U=0,~R+T\ne1$) in perfect TPQI, while the pure HOM dip requires full unitarity. 
\par
In conclusion, we have revealed new features of TPQI in the context of plasmonic structures which are intrinsically lossy. We demonstrated how TPQI can be used as a better probe of plasmonic resonances. Even the overlapping plasmonic resonances can be resolved by TPQI. We discussed the possibility of perfect TPQI and gave conditions for its occurence.
\par
One of the authors would like to thank Shourya and Kenan for help in the preparation of the code and the diagrams.


\begin{thebibliography}{99}

\bibitem{hom}
C. K. Hong, Z. Y. Ou, and L. Mandel, ``Measurement of Subpicosecond Time Intervals between Two Photons by Interference", \prl \textbf{59}, 2044 (1987).

\bibitem{gsa} G. S. Agarwal, \textit{Quantum optics}, (Cambridge University Press, Cambridge, 2013) Ch. 5.

\bibitem{dowling}
J. P. Dowling, ``Quantum optical metrology - the lowdown on high NooN states", Contemp. Phys. \textbf{49}, 125 (2008).

\bibitem{target1}
S. Tan, B. I. Erkmen, V. Giovannetti, S. Guha, S. Lloyd, L. Maccone,
S. Pirandola, and J. H. Shapiro,``Quantum Illumination with Gaussian States", \prl \textbf{101}, 253601 (2008).

\bibitem{qi1}
A. Kolkiran and G. S. Agarwal, ``Heisenberg limited Sagnac interferometry", \opex \textbf{15}, 6798 (2007).

\bibitem{qi2}
K-H. Luo, J. Wen, X-H Chen, Q. Liu, M. Xiao, and L-A. Wu,``Second-order Talbot effect with entangled photon pairs", \pra \textbf{80}, 043820 (2009).

\bibitem{woerdman}
E. Altewischer, M. P. van Exter and J. P. Woerdman, \nat \textbf{418}, 314 (2002).

\bibitem{zbinden}
S. Fasel, F. Robin, E. Moreno, D. Erni, N. Gisin, and H. Zbinden, ``Energy-Time Entanglement Preservation in Plasmon-Assisted Light Transmission", \prl \textbf{94}, 110501 (2005).

\bibitem{sahin}
G. Di Martino, Y. Sonnefraud, S. K\'ena-Cohen, M. Tame, S. K. \"Ozdemir, M. S. Kim and S. A. Maier, ``Quantum statistics of surface polaritons in metallic stripe waveguides", Nano Lett. \textbf{12}, 2504 (2012).

\bibitem{lukin}
A. V. Akimov, A. Mukherjee, C. L. Yu, D. E. Chang, A. S. Zibrov, P. R. Hemmer, H. Park and M. D. Lukin, ``Generation of single optical plasmons in metallic nanowires coupled to quantum dots", \nat \textbf{450}, 402 (2007).

\bibitem{zwiller} 
R. W. Heeres, L. P. Kouwenhoven and V. Zwiller, ``Quantum interference in plasmonic circuits", Nat Nanotech. \textbf{8}, 719 (2013).

\bibitem{cpa1}
W. Wan, Y. D. Chong, L. Ge, H. Noh, A. D. Stone and H. Cao, ``Time-reversed lasing and interferometric control of absorption", Science \textbf{331}, 889-892 (2011).

\bibitem{cpa2}
S. Dutta-Gupta, O. J. F. Martin, S. Dutta Gupta, G. S. Agarwal, ``Controllable coherent perfect absorption in a composite film'', \opex \textbf{20}, 1330 (2012). 

\bibitem{cpa3}
S. Dutta-Gupta, R. Deshmukh, A. V. Gopal, O. J. F. Martin, and S. Dutta Gupta, ``Coherent perfect absorption mediated anomalous reflection and refraction,'' \ol \textbf{37}, 4452 (2012).

\bibitem{sdgtf}
S. Dutta Gupta, ``Stratified Media for Novel Optics, Perfect Transmission and Perfect Coherent Absorption'', in \textit{Guided Wave Optics and Photonic Devices}, S. Bhadra, A. Ghatak, eds.(CRC Press, 2013), chapter 18, pp. 463-482.

\bibitem{pendry}
S. Tomita, T. Yokoyama, H. Yanagi, B. Wood, J. B. Pendry, M. Fujii and S. Hayashi, ``Resonant photon tunneling via surface plasmon polaritons through one-dimensional metal-dielectric metamaterials', \opex \textbf{16}, 9942 (2008).

\bibitem{sahin2}
J. Zhu, S. K. Ozdemir, Y-F. Xiao, L. Li, L. He, D-R. Chen and L. Y., ``On-chip single nanoparticle detection and sizing
by mode splitting in an ultrahigh-Q microresonator", Nat. Phot. \textbf{4}, 46 (2010).

\bibitem{sdgreview}
S. Dutta Gupta, ``Nonlinear optics of Stratified media,'' \textit{Progress in Optics}, E.Wolf, ed.(Elsevier Science, 1998), Vol. 38, pp.11-13.

\bibitem{johnson}
P. B. Johnson and W. Christy, ``Optical constants of the noble metals'', \prb\textbf{6}, 4370 (1972).


\end{thebibliography}

\newpage
\clearpage
\end{document}